\documentclass[12pt]{article}

\usepackage{amssymb,epsfig, here, tabularx, epsfig}
\newcommand{\C}{\mathbb{C}}
\newcommand{\He}{\mathbb{H}}
\newcommand{\HH}{{\mathcal{H}}}
\newcommand{\DD}{{\mathcal{D}}}
\newcommand{\JJ}{{\mathcal{J}}}
\newcommand{\AB}{{\mathcal{A}}}

\newcommand{\LL}{{\mathcal{L}}}

\newcommand{\MM}{{\mathcal{M}}}

\newcommand{\bequ}{\begin{equation}}
\newcommand{\dequ}{\end{equation}}
\newcommand{\bdis}{\begin{displaymath}}
\newcommand{\edis}{\end{displaymath}}

\begin{document}

\title{Left-Right Symmetric Models in Noncommutative Geometries?}
\author { Florian Girelli \\  \small Centre de Physique Th\'eorique, CNRS - Luminy, Case 907, 13288 Marseille Cedex, France
\\ \small girelli@cpt.univ-mrs.fr  }
\maketitle
\abstract{
Left-right symmetric models are analyzed in the
context of noncommutative geometry where we show
that spontaneous parity violation is ruled out... }  
\section{Introduction}
In the Standard Model of electro-weak and strong
forces, parity is broken explicitly by the choice of
inequivalent representations for left- and
right-handed fermions. Within the frame work of
Yang-Mills-Higgs models this is certainly an aesthetic
draw back that physicists have tried to correct by the
introduction of left-right symmetric models. These are
Yang-Mills-Higgs models where parity is broken
spontaneously together with the gauge symmetry.
However the price to pay for this aesthetic surgery is
high, especially on an aesthetic scale: the simplest
left-right symmetric model for electro-weak forces
has a $SU(2)_L\times SU(2)_R\times U(1)$ group and
four scalar representations transforming like two
$\underline 2_L\otimes
\underline 2_R$ representations, a
$\underline 3_L\otimes
\underline 1_R$ and a $\underline 1_L\otimes
\underline 3_R$. The Higgs potential contains some
twenty coupling constants \cite{binetruy}. For a tiny
class of Yang-Mills-Higgs models, noncommutative
geometry derives the Higgs mechanism, i.e. the scalar
representation and symmetry breaking Higgs
potential, from first principles. Parity violation is
crucial here in the sense that vector-like models are
not in this tiny class, the Standard Model, however,
with its explicit parity violation qualifies for
noncommutative geometry. Left-right symmetric
models are halfway in between vector-like models and
models with explicit parity breaking and its is
natural to ask whether they do qualify for
noncommutative geometry.
\\ In the Connes-Lott models \cite{connelott}, a first try was made but unfortunately did not qualify
\cite{is}. This was, however, before the setting of a precise notion of noncommutative geometry (Connes' Axioms \cite{axiom}) and the introduction of real structure which 
gave a very rigid structure to construction of finite spectral triples. A complete 
classification of these latter was made in \cite{kraj1}, enabling to construct 
some very general finite spectral triples in the framework of these axioms.
\\ Here we shall use this general classification  in order to study the most 
general LRS models, with the aim to check if they can be expected to be physical. 
\\ We shall first define the LRS models in the context of almost commutative geometries, by just specializing the usual LRS models 
to this approach. However,  Connes'Axioms imply a number of conditions that the model has to fulfill in order to be well-defined. 
In particular, we will see that having 2 groups acting respectively on the left and the right fermions will not be enough and we will need a third. 
It will appear also that the Poincar\'e duality cannot be satisfied.
\\ Secondly, we shall study some examples of LRS models as the updated (Spectral action \cite{spec} and respecting the Axioms) chiral electromagnetism. These examples will show that no parity breaking occurs and that, furthermore, the physical gauge bosons are either axial or vectorial. Finally, in the last part, we shall give a general proof of this result.
\section{Left Right Symmetric Models}
The LRS models are a particular type of YMH models. We recall that  a LRS model is based on  a product of 3
Lie groups, $G_L$, $G_R$ and  $G_V$.  $G_L$ and  $G_R$ acting respectively on $\psi _L , \psi _R$ and $G_V$   acts vectorially. 
LR symmetry means that $G_L$ and $G_R$ are isomorphic, $G_L \simeq G_R$ and that the representations on the left-handed fermions $\rho_L $ and on  right-handed 
fermions $\rho _R $ are identical, $\rho _L (G_L , G_V) = \rho _R (G_R , G_V):= \rho  (G_{L,R} , G_V)$.
\\ Another condition is needed. The Higgs vacuum $V_{vac}$ generates the fermionic masses as well the gauge bosons masses.
\bequ
<\psi , V_{vac} \psi > = < \psi _R, M \psi _L> + < \psi_L, M^* \psi _R>
\dequ
The matrix $M$ is in general not hermitian, and   we get the masses by a biunitary transformation $(U_L, U_R)$ acting such:
\bequ
U ^{-1} _R M U_L = \left( \begin{array}{ccc}
m_1 &&\\
& \ddots&\\
&&m_n
\end{array} \right)
\dequ
 However, in order to keep the symmetry between the Left and the Right fermions,
 we must  have only one  unitary transformation: $U_R=U_L :=U $. By definition, a  LRS model  has a Higgs vacuum (i.e. the fermionic mass matrix)
generated by a hermitian matrix $M$.
\\ The specialization to almost commutative models is then straightforward ( see also  a slightly different approach    in  \cite{saito}). The main object in almost commutative geometries is the 
associative algebra $\AB$ (say for example a real  one) equipped with a representation, noted $\pi$ (i.e. the fundamental one). The gauge bosons are the connections:
$\Omega ^1 (M,{ Lie U}(\AB))$ where $\MM$ is the space time manifold, and ${ Lie U}(\AB)$ the Lie algebra of the unitaries of $\AB$. The real structure $\JJ$ 
generates a bimodule structure on $\HH$, the Hilbert space of fermions. The group action on $\HH $ is then $\rho (U) \psi = \pi(U) \JJ \pi(U) \JJ ^{-1}\psi$ for $U \in U(\AB)$. 
To get the interactions of the gauge bosons with fermions, we have to go to the Lie algebra. Taking an infinitesimal element of $U(\AB)$, that is $u \in 
{\rm Lie U}(\AB)$,  the action of $LieU(\AB)$ on $\HH $ ( noted $\pi ^s$ for symmetrized representation) is then $\pi ^s (u)\psi =( \pi (u) + \JJ 
\pi (u) \JJ ^{-1})\psi$. 
\\ Applying this  to LRS model, we see  that   the finite 
algebra must be decomposed in $\AB _L , \AB _R$ and  $ \AB _V $ acting respectively on $\psi _L , \psi _R$ and vectorially, 
such that $\pi ^s( \AB_L , \AB _V) = \pi ^s(\AB _R , \AB _V)$ and $\AB _L \simeq \AB_R$. 
\\ The other condition about the Higgs vacuum (or the fermionic mass matrix) is not changed and can be taken as it is.
\\ This being set, let us check how Connes' axioms constrain our model. We shall deal with $\AB _L$ and $\AB _R$ being simple algebras as
these are the simplest to deal with and that the case of semi simple algebras can be easily recovered in the same way. 
\\ The fact that $\pi ^s _L = \pi ^s _R$ sets some very strong  conditions on the 
matrix  of multiplicities \cite{kraj1}.
First, let us consider the case  $\AB =\AB _L \oplus \AB _R$, that is with no $\AB _V$.
 $\AB _L , \AB _R$ acting respectively on the left and 
right-handed fermions,   are respectively affected by a sign  $\ominus$ and $\oplus$. 
The matrix of mutiplicities is therefore:
\begin{center}
\( \begin{array}{ccc}
  & \AB _L & \AB _R \\
\AB _L & \ominus &  \\
\AB _R& & \oplus
\end{array} \; \; \; \ominus, \oplus \textrm{ are resp. for } \pi ^s(\AB _L), \pi ^s(\AB _R) \)
\end{center}
In \cite{kraj1}, it was shown that to construct the most general Dirac operator, we have to draw vertical or horizontal lines between different signs, and that each line
was representing a non zero element in the Dirac operator.
\\ It is clear here that we can not draw any, therefore the Dirac operator is zero. This is not interesting us as we want to have nonzero fermionic masses. 
\\ Therefore we see that  we need a third algebra $\AB _V$ which would act on both left and right fermions. This is the same reason as in the Standard Model 
were we need the  $SU(3)$ color implemented to have a well-defined model.
\\ Bearing this in mind, we can easily establish the most general matrix of 
multiplicities:
\begin{center}
\( \begin{array}{cccc}
& \AB _L & \AB _R & \AB _V\\
\AB _L&\ominus _1 & & \ominus _2 \\
\AB _R&        &\oplus _3& \oplus _4 \\
\AB _V&\ominus _2 &  \oplus _4 & ? 
\end{array} \)
\end{center}     
$\ominus _1$ stands for $\pi ^s (\AB _L)$, $\ominus _3$ for $\pi ^s (\AB _R)$, $\ominus _2$ for $\pi ^s (\AB _L, \AB _V)$, $\ominus _4$ for $\pi ^s (\AB _R, \AB _V)$,  and ? for $\pi ^s (\AB _V)$
\\ ? can be $ \oplus $ or $\ominus $ or $void$ (here $\AB _V$ is supposed simple but nothing 
forbids us to take it semi-simple). 
\\ However if the action  $\AB _V$ was affected by a sign, it would mean that 
this algebra act  only on a particular fermion chirality. This would   
break the LR symmetry so the $?$ has to be $void$, that is the 
algebra  $\AB _V$ acts on  fermions of both chiralities, in the same way i.e. vectorially. 
\\ The signs in the matrix of multiplicities can be affected by relative numbers. Indeed, the reality condition implies that $\pi$ and $\JJ \pi \JJ ^{-1}$
 commute. Therefore, they can be decomposed into irreducible representations: $\pi (\AB )= \oplus_i M_{n_i} \otimes I_{m_i}$. Let us take $x_i \in M_{n_i} (\C )$ and
\begin{eqnarray}
\pi (\AB ) \ni \oplus _{ij} x_i \otimes I _{m_{ij}} \otimes I_{n_j} \nonumber \\
\JJ \pi (\AB ) \JJ ^{-1} \ni \oplus _{ij} I_{n_i} \otimes I _{m_{ij}}\otimes \overline{x_j}
\end{eqnarray}
The (symmetric) matrix of natural numbers $(m_{ij} )$ with each $m_{ij}$ affected by our previous sign is by definition the matrix of multiplicities \cite{kraj1}
 $\mu _{ij} = sign _{ij}\; m_{ij}$.
So now the most general matrix of multiplicities is:
\begin{center}
\( \mu = \left( \begin{array}{ccc}
\mu _{11} &         &  \mu _{13} \\
        &  \mu _{22}  &   \mu _{23} \\        
\mu _{31} &   \mu _{32} &      
\end{array} \right) = \begin{array}{ccc}
\ominus & & \ominus \\
        &\oplus & \oplus \\
\ominus &  \oplus  &  
\end{array} \)
\end{center}     
Obviously, we need $\mu _{11} = \mu _{22}$ and $\mu _{13} = \mu _{23}$ in order 
to have  $\pi ^s _L =\pi ^s_R$. But this is a problem as we can check that the matrix then
 has a zero determinant! This means that the Poincar\'e duality cannot be 
satisfied in this model \cite{kraj1}.  However we shall not dismiss  the LRS models for 
this "mathematical" reason, indeed we are going to see that also for a physical reason the model is not viable: parity remains unbroken. 
\\ The most general Dirac operator can be constructed and from it the Higgs fields. The computation of the Higgs potential then  depends on which scheme we 
are considering. However, our main interest concerns the Higgs vacuum which will give both fermionic  and gauge bosons masses. It is calculated
by minimizing the Higgs potential. In the case of Connes-Lott models,  we will get as minimum the Dirac, but in the case of the Spectral Action,
we do not know what we will get as the potential is a tricky polynomial to minimize.   
\\ However, in both cases the Higgs vacuum will be given in terms of the fermionic mass matrix $M$ and once again, 
 we must   have $M$  hermitian, in order to have the LR symmetry preserved.

\section{Examples}
\subsection{Chiral electromagnetism}
Let us study a simple example to see what is going on. The ancestor of the almost commutative Standard Model is the two 
sheeted model: $C^\infty (\MM) \otimes (\C _L \oplus \C _R )$ \cite{conne1}. This first  model showed that on the discrete part some 
spontaneous  symmetry breaking occurred, and therefore at the end, we  get one massive  and one massless gauge bosons. 
The massive acting axially and the latter  vectorially, this model was called  chiral electromagnetism. However  vectorial or axial interactions do not break parity. Indeed, in the mass Lagrangian $\LL _m = m^2 _{a} W_a 
^2 + m^2 _{\gamma} \gamma _v ^2$, to exchange the algebras $\C _L$ and $\C _R $ would change nothing as 
$W_a \rightarrow -W_a$ and $\gamma _v \rightarrow \gamma _v$.
\\ Chiral electromagnetism was introduced in the Connes-Lott scheme, and 
before the setting of the Axioms and the Spectral action \cite{axiom}\cite{spec}\cite{schuk}. 
\\ Here we are going to calculate the updated model (that is taking account of  Connes' Axioms and in the Spectral action scheme) 
and check explicitly that parity remains unbroken.
\\ As we saw in the last section,  we need to add another algebra to $\C _L$ and $\C _R $ in order to have the first order condition 
 verified.


 We choose to take the simplest one: $\C$. So our algebra is now $\C _L 
\oplus \C _R\oplus\C _V$ where $\C _V$ acts vectorially. Taking $(a, b, c) \in \C _L \oplus \C _R\oplus\C 
_V$, the chosen representation is (one family):
\begin{center} \( \left( \begin{array}{cccc}
a& & & \\
& b & & \\
& & \overline{c} & \\
& & & \overline{c} 
\end{array} \right) \; \; \textrm{acting on}  \left( \begin{array}{c}
e_L \\
e_R \\
\overline{e_L}\\
\overline{e_R}
\end{array}
\right) \in \HH \sim \C ^4.
\) \end{center}
Chirality and charge conjugaison are respectively given by:
\begin{center} \(\Gamma =  \left( \begin{array}{cccc}
-1&0 &0 &0 \\
0& 1 &0 &0 \\
0&0 & -1 &0 \\
0&0 &0 & 1 
\end{array} \right) \; \; \JJ = \left( \begin{array}{cccc}
0&0 &1 &0 \\
0& 0 &0 &1 \\
1&0 &0  &0 \\
0&1 &0 &0  
\end{array} \right) o\textrm{ complex conjugation}
\) \end{center}
Then the most general Dirac operator is 
\begin{center}
\( \DD = \left( \begin{array}{cccc}
0 & m &0 &0 \\
m^* & 0&0&0 \\
0&0&0&m* \\
0&0&m&0
\end{array} \right) \; \; \textrm{with} \; m\in \C. \)
\end{center}
The Higgs field is  parametrized by a complex field  $\phi \in \C$:
\begin{center}
\( \Phi = \left( \begin{array}{cccc}
0 & m \phi &0 &0 \\
m^*\phi ^* & 0&0&0 \\
0&0&0&0 \\
0&0&0&0
\end{array} \right) \; \; . \)
\end{center}
Using the Spectral action, we get the Higgs potential:
\bequ
V(\phi ) =- \textrm{tr} \Phi ^2 + \frac{1}{2} \textrm{tr} \Phi ^4 = -2 |m|^2 |\phi|^2 + |m|^4 |\phi|^4 .
\dequ
So, we get the Higgs vacuum given by:
\begin{center}
\( \Phi _{vac}= \left( \begin{array}{cccc}
0 & V_{vac} &0 &0 \\
V^*_{vac} & 0&0&0 \\
0&0&0&0 \\
0&0&0&0
\end{array} \right). \)
\end{center}
$V_{vac}$ is chosen   real. Next we want to  compute the mass matrix of 
the gauge bosons which is given by:
\begin{eqnarray}
\mathcal{M} = \textrm{tr} ([\pi (\AB ) , \Phi_{vac} + \JJ \Phi_{vac} \JJ ^{-1}] [\pi (\AB ) ,\Phi_{vac} + \JJ\Phi_{vac} \JJ ^{-1}]^*) \nonumber  \\
 = 4 \textrm{tr} ([\pi (\AB ) , \Phi_{vac} ] [\pi (\AB ) , \Phi_{vac} ]^*) \nonumber
\end{eqnarray}
Expanding in our case, we get:
\bequ
\mathcal{M} =(b^2 +a^2 ) V^2_{vac} -2abV^2_{vac}
\dequ
and this can be written as 
\begin{center}
\( (a,b) \left( \begin{array}{cc}
V^2_{vac}&-V^2_{vac} \\ 
-V^2_{vac}&V^2_{vac}
\end{array} \right) \left( \begin{array}{c}
a\\
b
\end{array}\right)\)
\end{center}
To know the mass of the physical bosons, we need to diagonalize this mass matrix. The rotation needed is \(\frac{1}{\sqrt{2}} 
\left( \begin{array}{cc}
1&1 \\
-1&1 
\end{array} \right) \) and we get the mass eigenvalues $2V^2_{vac}, 0$ associated respectively with the eigenvectors $W_A = 
a-b$ and $\gamma _V = a+b$. We get the announced result, that is an axial massive boson and  a vectorial massless boson, the photon. This means of course that parity is  unbroken.
\\  However this property is not linked to the fact that we took  commutative algebras. Indeed other models have been 
studied: $\He \oplus \He \oplus \C$ \cite{ralf} and $\He \oplus \He$ \cite{is}. The latter was studied before 
the Connes'Axioms setting. $\He \oplus \He \oplus\C$ is the updated version of $\He \oplus \He $: the $\C$ is there to 
make the model compatible with the Axioms. Both models were  done in the Connes-Lott scheme, and both gave that no parity 
breaking occurred and that we still had that the eigenvectors of the gauge bosons mass matrix were 
either vectorial or axial. Just as a last example, we can have a look of what goes on in the Spectral Action scheme with the same algebra, i.e. $\He \oplus \He \oplus\C$. (We recall that $\He$ is considered as a real algebra.)
\subsection{Case of $\He \oplus \He \oplus \C$
}
 We take the same representation as in \cite{ralf}: $(a, b, c) \in \He \oplus \He \oplus\C$ is represented as:
\begin{center} \( \left( \begin{array}{cccc}
a\otimes 1_2& & & \\
& b\otimes 1_2 & & \\
& & \overline{c} 1_2\otimes 1_2 & \\
& & & \overline{c}  1_2\otimes 1_2
\end{array} \right)\) \end{center}  acting on
\(  \left( \begin{array}{c}
\lambda_L \\
\nu_{\lambda L}\\
\lambda_R \\
\nu_{\lambda _R} \\
\overline\lambda_L{}\\\overline{\nu_{\lambda L}}\\\overline{\lambda_R}\\\overline{\nu_{\lambda _R}}
\end{array}
\right) \otimes \left(\begin{array}{cc}
 \lambda = {\rm electron} \\
\lambda ={\rm muon} \end{array} \right)\in \HH \sim \C ^{16}
\) 
\\ Chirality and charge conjugaison are respectively given by:
\begin{center} \(\Gamma =  \left( \begin{array}{cccc}
-1_4&0 &0 &0 \\
0& 1_4 &0 &0 \\
0&0 & -1_4 &0 \\
0&0 &0 &  1_4
\end{array} \right) ,\;\; \JJ = \left( \begin{array}{cccc}
0&0 &1_4 &0 \\
0& 0 &0 &1_4 \\
1_4&0 &0  &0 \\
0&1_4 &0 & 0 
\end{array} \right) o\textrm{ complex conjugatson}
\)
\end{center}
Then the  Dirac operator is taken to be \cite{ralf}:
\begin{center}
\( \DD = \left( \begin{array}{cccc}
0 & M &0 &0 \\
M^* & 0&0&0 \\
0&0&0&\overline{M} \\
0&0&\overline{M^*}&0
\end{array} \right) \; \; \textrm{with} \; M=p_1 \otimes M_e + p_2 \otimes M_{\nu} \in M_4(\C) \)
\end{center}
$p_i$ are given by $(p_i)_{kl} = \delta _{ki} \delta _{il}$ and $M_e, M_{\nu}\in M_2(\C)$. 
The Higgs is then constructed in the usual way.
\begin{center}
\( \Phi = \left( \begin{array}{cccc}
0 &  \phi _1 \otimes M_e + \phi _2 \otimes M_{\nu} &0 &0 \\
 \phi^* _1 \otimes M^*_e + \phi^* _2 \otimes M^*_{\nu} & 0&0&0 \\
0&0&0&0 \\
0&0&0&0 \end{array} \right)\) 
\end{center}
$\phi _i$  is defined by $\phi _i = h_i -p_i $ with $h_i \in M_2 (\C)$. One can show that the $\phi_i$'s are not independent \cite{ralf}: $\phi_2 = P_0 \overline{\phi_1}P_0 ^{-1}$ with $P_0= \left( \begin{array}{cc}
0& 1\\
-1 &0
\end{array} \right) $.
 The Higgs potential constructed from the Spectral action gives:
\bequ 
V(\Phi)= -\mu ^2 tr |\Phi|^2 + \lambda  tr |\Phi|^4
\dequ
In order to simplify the calculations we  take $M_{\nu}=0$, and then we want to minimize this potential. If we take $\Phi \Phi ^*$ as the variable then this minimum is obtained when $\phi$ is unitary, i.e. $\phi \in U(2)$ (modulo a renormalization factor $\kappa = \frac{\mu ^2 tr|M_e|^2}{2\lambda tr|M_e|^4 }$)
\bequ 
V_{vac}= \kappa u \otimes M_e \;\;\; u \in U(2)
\dequ
\begin{center} \( \Phi _{vac} =\kappa  \left( \begin{array}{cccc}
0& u \otimes M_e &0&0\\
u^* \otimes M^*_e &0&0&0\\
0&0&0&0\\
0&0&0&0
\end{array} \right) = \left( \begin{array}{cccc}
0&V_{vac}&0&0\\
V^*_{vac} &0&0&0\\
0&0&0&0\\
0&0&0&0
\end{array} \right)\)
\end{center}In order to keep the LR symmetry, we have to choose $V_{vac}$ to be hermitian, and fermionic masses are obtained by diagonalizing  $V_{vac}$.
\\ This is a striking result, indeed the Dirac operator is not a minimum of the Higgs  potential, as one could expect. To the best of my knowledge, this is the first example of this type. 

We consider now the mass matrix for the gauge bosons:
\bequ \MM _b = \textrm{tr} 
([\pi (\AB ) , \Phi_{vac} ] [\pi (\AB ) , \Phi_{vac} ]^*)
\dequ
 The gauge bosons (elements of the complexification of the  Lie algebra) are respectively for $\AB _L$ and $\AB _R$ $a_{\mu}=\left( \begin{array}{cc} 
A^3_{\mu} & A^+_{\mu} \\
A^-_{\mu}  & -A^3_{\mu} \end{array} \right)$ and $  b_{\mu} =\left( \begin{array}{cc} 
B^3_{\mu} & B^+_{\mu} \\
B^-_{\mu}  & -B^3_{\mu} \end{array} \right)$ and they can be taken  hermitian. So we get for the mass matrix:
\bequ
\MM _{b} =\textrm{tr}( V_{vac}V^*_{vac}( a^2_{\mu}\otimes 1_2)+ V^*_{vac}V_{vac}( b^2_{\mu}\otimes 1_2) ) - 2 ((a_{\mu}\otimes 1_2) V_{vac}(b_{\mu}\otimes 1_2) V^*_{vac})).
\dequ
First we can take the  simple value for $V_{vac}$, 
 $u=1_2$.
Straightforward calculations (diagonalization of  $\MM _{b}$) show that there are 3 massless bosons, all of them vectorial,  one being neutral, the two others charged. All  massive bosons are  axial and  have the same mass: $m_Z = m_{W^{\pm}} $. Another way of counting  massless bosons, is to calculate the little group of the potential: It is given by the $u_L , \; u_R \in SU(2)$ such that $u_L u u^{-1}_R = u$. For the simple vacuum expectation value  $u=1_2$, the little group is $SU(2)$ with $u_L=u_R$ and there are 3 massless bosons.
\\ If now  we take $u$ to be a general element of $U(2)$, the mass matrix is more complicated to calculate and  we compute then  the little group. Indeed a general  $u\in U(2)$ is diagonalizable by a biunitary transformation $P_L, P_R \in U(2)$ such that $P_LuP_R ^{-1}=1_2$. Then the equation $u_L u u^{-1}_R = u$ becomes 
\begin{eqnarray}
u_L P^{-1}_L P_R u_R ^{-1} = P_L ^{-1} P_R \nonumber \\
 P_L u_L P^{-1}_L P_R u_R ^{-1} P_R ^{-1}:= u'_L u ^{'-1}_R =1_2. \nonumber
\end{eqnarray}
Therefore $u_R = ( P_R ^{-1} P_L)u_L  ( P_R ^{-1} P_L) ^{-1}$ and if $u_L$ is of determinant one, so is $u_R$. The little group is $SU(2)$ again.
 One checks also that the massless bosons are still vectorial and the massive ones, all of the same mass,  are axial. 
\\ Parity is still not broken.
\\ Let us compare our  $\He \oplus \He \oplus \C$ example in the Spectral action scheme to the  $\He \oplus \He \oplus \C$ example in the  Connes-Lott scheme  \cite{ralf}:
\begin{itemize}
\item Firstly we saw that starting with a Dirac operator  with rank one in the isospin sector, we ended up with a fermionic mass matrix of rank 2 (maximal rank) in the isospin sector, whereas in the Connes-Lott scheme, the initial Dirac operator used to construct 1-forms is also the minimum of the Higgs potential. \cite{ralf} 
\item Secondly, this special  feature implies a difference between gauge bosons masses: instead of the  unique massless neutral boson in the Connes-Lott scheme  \cite{ralf}, we get 2 more charged massless bosons. In terms of little groups, Connes-Lott has  $U(1)$ as little group,  whereas the Spectral action has  $SU(2)$.
\end{itemize}  
Let us note that the Standard Model does not exhibit these differences. Indeed, here the Dirac operator also minimizes the Higgs potential derived from the Spectral action, and the little groups coincide in both schemes. This is another remarkable feature of the Standard Model in noncommutative geometry. 

Instead of calculating all  masses by hand, there is a quicker way to  show that  parity is  not broken. Indeed,  we do not even need to calculate the minimum explicitly: let us take it as 
\begin{center} \( \Phi _{vac}  = \left( \begin{array}{cccc}
0&V_{vac}&0&0\\
V^*_{vac} &0&0&0\\
0&0&0&0\\
0&0&0&0
\end{array} \right)\)
\end{center}
We consider now the mass matrix for the gauge bosons: $\MM = \textrm{tr} ( V_{vac}V^*_{vac}( a^2_{\mu}\otimes 1_2)+ V^*_{vac}V_{vac}( b^2_{\mu}\otimes 1_2) ) - 2 ((a_{\mu}\otimes 1_2) V_{vac}(b_{\mu}\otimes 1_2) V^*_{vac})).$
Here, we can   notice  that by interchanging $a_{\mu}$ and $b_{\mu}$ the matrix $\MM$ is not modified. This switching is just the action of  the matrix $\Sigma $ which can be written  in the chosen basis as $\Sigma = \left( \begin{array}{cc} 
0&1 \\
1&0 \end{array}\right) ^{\oplus 3}$. We have of course that $\Sigma = \Sigma ^{-1}=  \Sigma^{t}$. Therefore  
\bequ
\MM = \Sigma \MM \Sigma .
\dequ
We can see then that $\Sigma$ and $\MM$ are commuting and so diagonalized in the same eigenbasis. By construction,  $\Sigma$ has eigenvalues $-1, 1$. Let us take $W _{\mu}$ an eigenvector of $\MM$ with $m$ as eigenvalue: $ \MM W _{\mu} = m  W _{\mu}$. It is also an eigenvector of $\Sigma$ which means therefore that $\Sigma W _{\mu} = \pm W _{\mu}$. This last equation is just asserting that $W_{\mu}$ is either axial or vectorial and  parity remains unbroken!  
\\ These results suggest us  the following theorem.
\section{Theorem}
{\it In  almost commutative YMH models, parity cannot be spontaneously 
broken,  moreover the physical gauge bosons, mass eigenstates,  are always vectorial or axial}.
\\  Let us prove this.
\subsection*{ Proof}
 Let us take the general case of $\AB = \AB _L \oplus \AB _R \oplus \AB _V$. We always have the possibility to reorder the Hilbert basis such that  
$\pi _t (\AB _{L,R} , \AB _V) = \pi(\AB _{L,R}) \oplus 
\pi -V(\AB _V)$. The mass matrix of the gauge bosons is then given:
\bequ
\mathcal{M} =tr ([\pi _t (\AB ) , V_{vac} ] [\pi _t (\AB ) , V_{vac} ]^*)= tr((\pi(\AB _L ) M - M\pi (\AB _R ) ) (\pi(\AB _L ) M -M \pi (\AB _R))^*)
\dequ
This formula holds  in any scheme, Connes-Lott or Spectral action, this is why our proof is independent of the used scheme. 
We note also that $\AB _V$ does not occur in our calculations: vector-like groups are never broken in almost commutative geometries.
\\ The matrix $\MM$ is by construction hermitian and has  positive (or zero) eigenvalues, indeed $\MM$ represents  a positive bilinear form.
\\ In order to check if  parity breaking can occur, we have to see what happens upon  exchange of $\AB _{L}$ and $\AB 
_{R}$, that is under the permutation of left and  right gauge bosons. The  permutation, noted $\Sigma$,  verifies obviously:
\bequ
\Sigma=\overline{\Sigma }=\Sigma ^t=\Sigma ^{-1}
\dequ
and its eigenvalues are  $1$ and $-1$.
\\ The exchange of $\AB _L$ and $\AB _R$ is thus given by:
\bequ
\Sigma \MM \Sigma = tr((\pi(\AB _R ) M - M \pi (\AB _L ) ) (\pi(\AB _R ) M - M \pi (\AB _L))^*)
\dequ
Working on it:
\bequ
\MM = tr((M^* \pi ^*(\AB _R )-\pi ^*(\AB _L )M)^*(M ^*(\pi ^*(\AB _R )-\pi ^*(\AB _L)M^*)
\dequ
Recall now that $M$ is hermitian, and that $\pi (\AB _{L,R})$ is anti hermitian since the gauge bosons belong to the Lie algebra $Lie U(\AB)$, so we get:
\begin{eqnarray}
\Sigma \MM \Sigma & = & tr((\pi(\AB _R ) M - M \pi (\AB _L ) ) (\pi(\AB _R ) M - M \pi (\AB _L))^*) \nonumber \\
                &    = & \MM
\end{eqnarray}
Since  $\Sigma = \Sigma ^{-1} $, we see that $\Sigma $ and $\MM$ commute and  can be diagonalized in the same eigenbasis. 
Let $W$ be such a simultaneous  eigenvector of $\MM$: 
\bequ
\Sigma W = \pm W.
\dequ
This last expression just expresses that $W$ is either vectorial or axial! Of course, by construction, we have as many vectorial bosons as axial ones. 
\\ It is obvious now that in any case there can be no parity breaking!

\section{Conclusion}

In conclusion, explicit parity violation is a must in
noncommutative geometry. On the mathematical side,
this follows from Poincar\'e duality. On the more
physical side, it follows from the above theorem which
rules out spontaneous parity breaking. There is an
impressing list of intricate features of the
Standard Model, that remain completely ad-hoc in the
context of Yang-Mills-Higgs models and that are
unavoidable in noncommutative geometry:
\begin{itemize}\item
the gauge group is non-simple,
\item
fermions transform
according to fundamental or trivial representations
under isospin and color,
\item
strong forces couple vectorially,
\item
color is unbroken,
\item
isospin is broken spontaneously by one
 doublet of scalars,
\item
 the gauge group is reduced by
${\mathbb Z}_2$.
\end{itemize}
We may now add to this list:
\begin{itemize}
\item Parity violation is explicit.
\end{itemize}
{ \bf Acknowledgements}
\\ T. Sch\"ucker is warmly thanked for suggesting the subject of this article as well as for his many advices.  B. Iochum and T. Krajewski are also  very much thanked for all their comments and advices.

\end{document}